\documentclass[pdflatex,mathphys]{sty}

\usepackage{upgreek}

\jyear{2023}%

\unnumbered

\begin{document}

\title[Real-time photonic RF interference solver]{A system-on-chip microwave photonic processor solves dynamic RF interference in real time with picosecond latency}

\author*[1]{\fnm{Weipeng} \sur{Zhang}}\email{weipengz@princeton.edu}
\author[1]{\fnm{Joshua C.} \sur{Lederman}}\email{joshuacl@princeton.edu}
\author[2]{\fnm{Thomas} \sur{Ferreira de Lima}}\email{thomas@nec-labs.com}
\author[1]{\fnm{Jiawei} \sur{Zhang}}\email{jz8939@princeton.edu}
\author[1]{\fnm{Simon} \sur{Bilodeau}}\email{sbilodeau@princeton.edu}
\author[1]{\fnm{Leila} \sur{Hudson}}\email{lshudson@princeton.edu}
\author[3]{\fnm{Alexander} \sur{Tait}}\email{alex.tait@queensu.ca}
\author[4]{\fnm{Bhavin J.} \sur{Shastri}}\email{shastri@ieee.org}
\author*[1]{\fnm{Paul R.} \sur{Prucnal}}\email{prucnal@princeton.edu}

\affil*[1]{\orgdiv{Department of Electrical and Computer Engineering}, \orgname{Princeton University}, \orgaddress{\city{Princeton}, \postcode{08544}, \state{New Jersey}, \country{USA}}}

\affil[2]{\orgname{NEC Laboratories America}, \orgaddress{\city{Princeton}, \postcode{08540}, \state{New Jersey}, \country{USA}}}

\affil[3]{\orgdiv{Department of Electrical and Computer Engineering}, \orgname{Queen’s University}, \orgaddress{\city{Kingston}, \postcode{K7L 3N6}, \state{Ontario}, \country{Canada}}}

\affil[4]{\orgdiv{Department of Physics, Engineering Physics and Astronomy}, \orgname{Queen’s University}, \orgaddress{\city{Kingston}, \postcode{K7L 3N6}, \state{Ontario}, \country{Canada}}}

\abstract{
Radio-frequency interference is a growing concern as wireless technology advances, with potentially life-threatening consequences like interference between radar altimeters and 5G cellular networks. Mobile transceivers mix signals with varying ratios over time, posing challenges for conventional digital signal processing (DSP) due to its high latency. These challenges will worsen as future wireless technologies adopt higher carrier frequencies and data rates. However, conventional DSPs, already on the brink of their clock frequency limit, are expected to offer only marginal speed advancements. This paper introduces a photonic processor to address dynamic interference through blind source separation (BSS). Our system-on-chip processor employs a fully integrated photonic signal pathway in the analogue domain, enabling rapid demixing of received mixtures and recovering the signal-of-interest in under 15 picoseconds. This reduction in latency surpasses electronic counterparts by more than three orders of magnitude. To complement the photonic processor, electronic peripherals based on field-programmable gate array (FPGA) assess the effectiveness of demixing and continuously update demixing weights at a rate of up to 305 Hz. This compact setup features precise dithering weight control, impedance-controlled circuit board and optical fibre packaging, suitable for handheld and mobile scenarios. We experimentally demonstrate the processor's ability to suppress transmission errors and maintain signal-to-noise ratios in two scenarios, radar altimeters and mobile communications. This work pioneers the real-time adaptability of integrated silicon photonics, enabling online learning and weight adjustments, and showcasing practical operational applications for photonic processing.
}


\maketitle
\section{Introduction}

\begin{figure}[ht!]
\centering
\includegraphics[width=.8\linewidth]{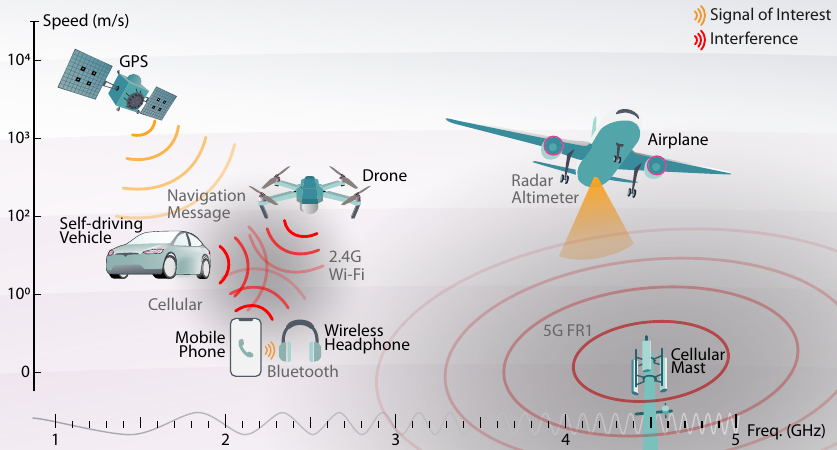}
\centering
\caption{\textbf{Scenarios of dynamic RF interference.} Red curves stand for interference signals. Orange-coloured shadings denote signals of interest. One example of spectrum resource congestion is the interference within industrial, scientific and medical (ISM) bands near 2.4 GHz, occupied by mobile cellular, Bluetooth, Wi-Fi, and amateur radios. In this scenario, the moving speeds of involved users, such as mobile phones, wireless headphones, ground vehicles and drones, span several orders of magnitude. The second depicted example of interference is between the aeronautical radio-navigation band (4.2 - 4.4 GHz) and the fifth-generation cellular band (below 4.1 GHz and above 4.5 GHz). This interference is relevant to all types of aircraft using onboard radar altimeters to measure their altitude above the terrain. Extended usage of radar altimeters affects many critical flight functionalities, such as terrain awareness, aircraft collision avoidance, and automatic flight control.
}
\label{fig:vision}
\end{figure}

Radar altimeters are the sole indicators of altitude above a terrain. Spectrally adjacent 5G cellular bands pose significant risks of jamming altimeters \cite{sanders2022measurements,son2020interference} and impacting flight landing and take-off. As wireless technology expands in frequency coverage \cite{dang2020should,shafique2020internet} and utilises spatial multiplexing \cite{kutty2015beamforming,larsson2014massive} (see Fig. \ref{fig:vision}), similar detrimental radio-frequency (RF) interference becomes a pressing issue. To address this interference, RF front-ends with exceptionally low latency are crucial for industries like transportation, healthcare, and the military, where the timeliness of transmitted messages is critical \cite{kim2018massive,bennis2018ultrareliable,ashraf2017towards,mozaffari2016unmanned,simsek20165g}. Future generations of wireless technologies will impose even more stringent latency requirements on RF front-ends due to increased data rate, carrier frequency, and user count \cite{cisco2020cisco}. Additionally, challenges arise from the physical movement of transceivers, resulting in time-variant mixing ratios between interference and signal-of-interest (SOI). This necessitates real-time adaptability in mobile wireless receivers to handle fluctuating interference, particularly when it carries safety-to-life critical information for navigation and autonomous driving \cite{roos2019radar,morales2019survey,dempster2016interference}, such as aircraft and ground vehicles.

Photonic integrated circuits (PIC) can handle broadband information by upconverting radio frequencies to optical frequencies in the hundreds of terahertz range. In contrast, analogue RF components have limitations in supporting wide bandwidths, while digital electronics inherently have clock-dependent performance and processing delays due to analogue-digital interconversions. As future wireless technologies adopt higher frequencies, these issues are expected to worsen. However, PICs overcome these limitations by offering ultra-low latency through direct analogue processing \cite{yang2022multi,daulay2022ultrahigh,huang2021silicon,shastri2021photonics,hu2020reconfigurable,marpaung2019integrated,perez2017multipurpose,wang2015reconfigurable,ghelfi2014fully}. They have demonstrated reduced latency of tens of nanoseconds, even when implementing electrical and optical conversions off-chip \cite{zhang2022broadband,lederman2023real}. By integrating on-chip modulators and photodetectors, a fully integrated photonic RF signal pathway holds the potential for significant latency reductions, surpassing the capabilities of existing electronic chips and ensuring technological readiness for the future.

Despite its potential, such a ``system-on-chip'' for microwave processing is still in its early stages, facing significant challenges related to design, control, and packaging complexities. Previous demonstrations have rarely achieved a compact footprint while fully realising the potential of low latency \cite{bogaerts2020programmable,perez2020multipurpose}. Additionally, current PICs lack complete co-packaged electronic peripherals required for real-time performance. PICs have primarily operated with pre-trained static weights or weight sequences and rarely demonstrated the ability to adjust weights ``on the fly,'' i.e., online based on learning from the processed output. Previous setups relied on bulky devices for signal digitisation, analysis, and PIC control, often utilising oscilloscopes, computers, and current/voltage sources \cite{zhang2022broadband,ghelfi2014fully}. Offline processing, inter-device communications, and single-threaded execution led to slow control, and undesirable size, weight and power (SWaP) metrics. As a result of these factors, the non-real-time operation and unfavourable form factor of PIC-based systems make them unsuitable for real-world deployment and mobile applications unless the issues above are addressed and resolved.

Another essential requirement is an interference suppression mechanism that is easy to implement and capable of real-time execution. Radio sources are typically distinguished based on prior knowledge of their characteristics. For example, spectral filtering can differentiate sources by their centre frequencies; adaptive beamforming can block out signals based on their direction of arrival \cite{kutty2015beamforming, li2005robust}; and cognitive radio can proactively coordinate spectral resources for multiple users \cite{akyildiz2006next,haykin2005cognitive}. However, acquiring prior knowledge poses practical challenges, including independent operations of service providers, considerations for privacy protection \cite{fragkiadakis2012survey,clancy2008security,acquisti2006there}, and difficulties arising from the randomness and burstiness of transmitted data. Even without considering these obstacles, real-time operability can still be compromised by delays in obtaining channel state information and training for interference prediction \cite{selvi2018markov,biguesh2006training}. Fortunately, blind source separation (BSS) \cite{naik2014blind,choi2005blind} eliminates the need for prior knowledge by recovering signals from mixtures using simple statistical analysis of demixed outputs. This enables unmatched compatibility with diverse RF environments, regardless of spectrum density and mobility. Furthermore, BSS performs linear demixing, which can be fully implemented on photonic devices such as microring resonator (MRR) weight banks \cite{de2022design,zhang2022silicon,tait2016microring}, meshes of Mach-Zehnder modulators (MZM) \cite{perez2020multipurpose,perez2017multipurpose,clements2016optimal} and crossbar arrays of phase-change materials (PCM) \cite{fang2022ultra, feldmann2021parallel}. These photonic processors have demonstrated exceptional energy efficiency \cite{tait2022quantifying} and instantaneous bandwidth \cite{marpaung2019integrated,khan2010ultrabroad} that eliminates the need for frequency-switching mechanisms, facilitating the development of future wireless technologies. Additionally, photonic processors achieve accurate analogue demixing with a high dynamic range not limited by fixed digital resolution. Competitive bit precision can be achieved through control methods, such as the recently proposed dithering technique \cite{zhang2022silicon}. Silicon photonics offers a viable realisation of these processors, allowing chip-scale integration and simplified interfacing compatible with complementary metal-oxide-semiconductors (CMOS).

Here, we demonstrate a system-on-chip silicon photonic BSS setup co-packaged with complete electronic peripheral circuitry, effectively addressing dynamic RF interference issues. The photonic signal pathway is fully integrated on-chip, incorporating modulators, MRR weight banks, and photodetectors, resulting in a processing latency of less than 15 picoseconds. Furthermore, full peripheral functionalities are implemented within a single field-programmable gate array (FPGA) chip, significantly reducing the setup from several benches (tens of square meters) to just a few circuit boards. Within the FPGA, integrated analogue-to-digital converters (ADCs) and programmable logic gates perform high-throughput statistical analysis. Moreover, a built-in ARM processor handles high-level BSS algorithms and PIC control. This architecture benefits from accelerated processing and minimised inter-device communications, enabling real-time execution at a refresh rate of 305 Hz. We also program the FPGA to incorporate the dithering method for PIC control, facilitating highly accurate demixing of the signal mixtures. Alongside various engineering solutions, such as stable optical coupling and compact circuit board design, we package the entire photonic BSS setup as a palm-sized standalone device, ensuring portability and robustness for field deployment. We subject this device to testing and validation in two experimental emulations of dynamic interference scenarios, including mobile communications and radar altimeters. The obtained transmissions demonstrate error-free operation over 800 bits and maintain signal-to-noise ratios of over 15 dB throughout the testing process. Additionally, this consistent success in recovering SOI in the presence of moving transmitters reveals its potential for solving real-world applications.

\section{Results}
\subsection{Low-latency photonic processing}

Under RF interference, the received signals are mixtures of the desired SOI and unwanted interference, which can be assumed to be uncorrelated. According to the central limit theorem, when two uncorrelated signals are mixed \cite{barany2007central}, they become more Gaussian-like. This principle provides a viable method for extracting original signals by maximising the non-Gaussianity of demixed output. Some prerequisites are that the mixtures should result from different mixing ratios and that the number of mixtures is equal to or greater than the number of original signals. Our proposed BSS algorithm utilises kurtosis as a statistical indicator of non-gaussianity and employs an optimisation procedure to determine the demixing weights that result in the optimal kurtosis \cite{zhang2022broadband,huang2022high,ma2020blind}. Kurtosis ($\kappa$) is the fourth standardised moment, calculated from the fourth central moment ($\mu_4$) and the standard deviation ($\delta$), as $\kappa = \mu_4/\delta^4$. A commonly used optimisation method is the Nelder-Mead (NM), as it does not require the objective function to be differentiable and can handle noisy problems. We use a simplex composing $n+1$ sets of demixing weights for $n$ total input mixtures, and NM finds the optimal by iteratively reflecting, expanding, contracting, and shrinking the simplex.

\begin{figure}[ht!]
\centering
\includegraphics[width=.99\linewidth]{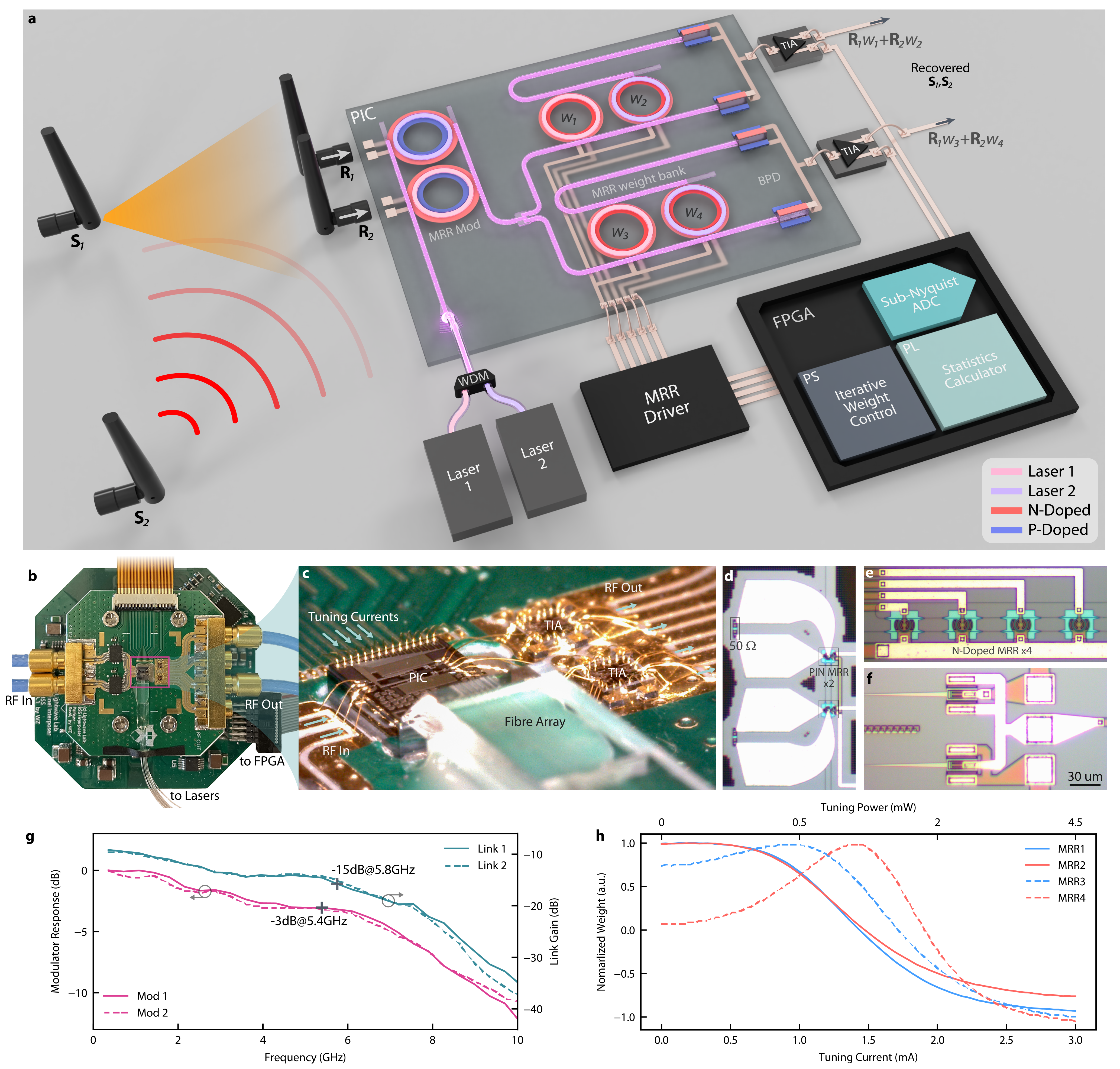}
\centering
\caption{\textbf{Photonic processor setup.} \textbf{a} Schematic of the photonic processor. PIC, photonic integrated circuit. MRR, microring resonator. Mod, modulator. BPD, balanced photodetector. TIA, transimpedance amplifier. FPGA, field-programmable gate array. Signal pathway starts at MRR Mod and ends at BPD, composing an on-chip waveguide with a length of 1.6mm, which has an index of refraction near 2.44. The latency is, accordingly, the light propagation time, which is about 15 ps. \textbf{b} Packaged palm-sized photonic processor. The setup comprises two printed circuited boards piled together and connected by a ribbon cable. The top board mounts the PIC is impedance engineered for handling high-frequency signals. The bottom board is populated with multichannel digital-to-analogue converters (DACs), providing tuning currents and biasing voltages for components on the PIC. Complete setup is shown in Supplementary Fig. 1. \textbf{c} Zoomed-in view of the packaged PIC, TIAs, and fibre array. \textbf{d-f} Micrographs of MRR modulators, MRR weight bank, and BPD, respectively. The ring modulators are of PN type and are slightly reversed biased by 0.55 V. This work uses two (the leftmost and the rightmost ones) of the four MRRs. The ring waveguide of each MRR is n-type doped to act as a heater for resonance tuning. The BPD has responsibility for about 1 A per watt and is biased about 1.1 V. \textbf{g} Modulator response and full link loss of the integrated signal pathway. Cross marks indicate a 3-dB bandwidth of 5.4 GHz and -15 dB link loss of up to 5.8 GHz. \textbf{h} Tuning current for thermal weight tuning of the two employed MRRs.
}
\label{fig:schematic}
\end{figure}

In all fairness, the computational scale for this interference cancellation is modest, just a 2x2 matrix operation for the case of one interference and one SOI. However, it necessitates exceptional bandwidth, latency, and power consumption performance. Conventional digital electronics struggle to meet these requirements, but our proposed photonic approach to signal processing provides a comprehensive solution, as elaborated upon.

Our photonic solution for RF interference uses a cohesive combination of a photonic processor for linear demixing and digital peripheral circuitry for statistical analysis and PIC control. Employing a fully integrated photonic signal pathway, as shown in Fig. \ref{fig:schematic}a, allows for direct demixing in the analogue domain determined by the time of flight, resulting in a latency of fewer than 15 picoseconds. Even considering potential latency added by the RC response of the modulator, the delay of the TIAs, and transmission time associated with the board traces, the overall latency (from RF In to RF Out as shown in Fig. \ref{fig:schematic}b) does not exceed 200 picoseconds. In contrast, digital electronic counterparts using a clock frequency of 100 MHz require 240 nanoseconds just for linear demixing using Wallace tree multipliers (a standard binary integer multiplier in hardware) \cite{wallace1964suggestion}, not including additional latency for analogue-to-digital and digital-to-analogue conversions. The calculation uses similar specifications as the photonic processor, with data representation of 16-bit integers and a weighting resolution of 9 bits. This reduced latency in suppressing interference is crucial in situations requiring immediate action for human safety, such as in military and transportation contexts.

On this chip, two PN-type MRRs modulate the input RF signals onto light from two lasers (Pure-Photonics, PPCL500) with wavelengths tuned to align with the corresponding resonance frequencies of the MRR modulators. These modulators support broad bandwidth (shown in Fig. \ref{fig:schematic}(g)) using the free-carrier dispersion effect of the reverse-biased PN junction of each MRR. The 3-dB bandwidth extends over 5.4 GHz and maintains a usable modulation efficiency ($<$10 dB roll-off) all the way up to 9.7 GHz. Following the modulation, the lightwaves are evenly divided, with each segment undergoing separate processing by a corresponding MRR weight bank, as depicted in Fig. \ref{fig:schematic}e. These weight banks perform individual weighting of RF signals carried by laser lights of different wavelengths, permitting scalability via wavelength-division-multiplexing (WDM) \cite{tait2014broadcast}. This technique has also been proved with an exceptional performance of over 19.2 GHz bandwidth \cite{zhang2022broadband} and 9 bits weighting accuracy \cite{zhang2022silicon}. The MRR weight bank further offers efficient thermal tuning capabilities. Transitioning from the most negative to the most positive weight requires less than 2.5 mA of current (as displayed in Fig. \ref{fig:schematic}h), consuming energy of less than 4.5 mW for each MRR and 18 mW for performing the 2x2 matrix operation. In contrast, a typical electrical processor requires 172 mW, given an energy consumption of 0.43 pJ per weighting \cite{jouppi2017datacenter} and considering a processing speed of 100 GS per second.

Next, a balanced photodetector (BPD), in conjunction with each MRR weight bank, calculates the differential in optical power between the drop and thru waveguide, converting the signals back to the electrical domain. Each output photocurrent is forwarded to a transimpedance amplifier (TIA, HMC7590, Analog Devices), which provides two duplicate amplified outputs. We feed one to an FPGA for the following statistical analysis while the other serves as the system output. Each set of a weight bank, a BPD, and a TIA compose a signal processing link. Two such links function concurrently, offering two individually weighted additions to the input signals that enable the recovery of both original signals in parallel. The overall link losses remain less than 15 dB over 5.8 GHz.

\subsection{Real-time operability for dynamic interference}

In a dynamic RF interference environment involving moving transceivers, extracting the SOI requires constantly updating demixing weights to keep up with time-variable mixing ratios. To achieve this, we co-package a photonic processor with a high-speed digital peripheral based on an FPGA board (Xilinx, RFSoC4x2. See Supplementary Fig. 3). The FPGA integrates multiple functional tiles, including an analogue-to-digital converter (ADC), programmable logic (PL), and an ARM processing system (PS) into a single chip die. The ADC digitises signals at a sampling rate of 4.915 GSPS and a length of $2^{15}$ samples for each frame. The PL processes sampled data and calculates kurtosis in a pipeline at the synchronised rate. The PS runs an application that performs an NM optimisation, determines the weights for each iteration, and commands the MRR driver via an express serial peripheral interface (SPI). This architecture speeds up the digital back-end by replacing old inter-device communication with rapid inner-chip communication and high parallelism statistical calculation, taking the place of single-thread CPU processing. In this new implementation, the time required for communication, statistical analysis, and MRR tuning is about 0.3, 2, and 1 milliseconds, respectively, achieving an accelerated update rate of 305 Hz. Solving two input mixtures, as demonstrated in this work, the NM algorithm requires a mere ten iterations (33 ms) for convergence to correct demixing weights from random initial guesses within. Thanks to the high update rate, it can track the optimal weights during each iteration when the mixing matrix varies over time. More details about the NM algorithm and the FPGA design can be found in Methods and Supplementary Information, respectively.

An accurate kurtosis measurement can use a sampling frequency much lower than the signal frequency, namely sub-Nyquist sampling, as shown in earlier works \cite{ma2020photonic, tait2018blind}. A lower sampling rate of FPGA does not affect the overall processing bandwidth and, in turn, has benefits in reduced power consumption, consuming less than 10 W for the complete system. In terms of SWaP, this co-packaging additionally reduces size and weight. A couple of printed circuit boards (PCBs) substitute several bulky devices in old setups (such as scopes, computers, and source measurement units). We have epoxy-glued a reflective fibre array for optical coupling to ensure efficient and mechanically stable coupling with a grating coupler in a low-profile setup. As shown in Fig. \ref{fig:schematic}b, the packaged setup has a tiny form factor of 40 mm by 30 mm, making it mobile-friendly as opposed to previously demonstrated setups that were only implementable in laboratory benches.

A high update rate is necessary to address time-variable mixing ratios in dynamic interference. To verify this, consider two transmitters (an SOI and an interferer) are received by a 2x2 multiple-input and multiple-output (MIMO) antenna, whose two orthogonal polarisation filters give two different mixtures of the original signals. Taking into account the path loss and denoting positions of the antennas as $\textbf{p}_{\textrm{tx1}}(t)$, $\textbf{p}_{\textrm{tx2}}(t)$, and $\textbf{p}_{\textrm{rx}}(t)$, the amplitude of the $i$th received mixture can be denoted as in Eq. \ref{eq:received_mixture_i}.

\begin{equation}
\begin{aligned}
	\textbf{R}_i(t) &= \frac{\lambda_{\textrm{tx1}}}{4\pi \|\textbf{p}_{\textrm{tx1}}(t)-\textbf{p}_{\textrm{rx}}(t)\| }G_{\textrm{tx1},i}(t)\textbf{S}_{\textrm{tx1}}(t)\\
     &+\frac{\lambda_{\textrm{tx2}}}{4\pi \|\textbf{p}_{\textrm{tx2}}(t)-\textbf{p}_{\textrm{rx}}(t)\|}G_{\textrm{tx2},i}(t)\textbf{S}_{\textrm{tx2}}(t),\,i=1,2
\end{aligned}
\label{eq:received_mixture_i}
\end{equation}
where $\textbf{S}_{\textrm{tx1}}(t)$ and $\textbf{S}_{\textrm{tx2}}(t)$ are the original signals. $G_{\textrm{tx1},i}(t)$ and $G_{\textrm{tx2},i}(t)$ represent the total antenna-related gains concerning the $i$th polarisation slants. These two gains are functions of time and are determined by the positions, orientations, and radiation patterns of antennas. $\lambda_{\textrm{tx1}}$ and $\lambda_{\textrm{tx2}}$ are the wavelengths of two original signals. Since the interference uses a frequency spectrum adjacent to the SOI, we can assume $\lambda_{\textrm{tx1}} \approx \lambda_{\textrm{tx2}} = \lambda$. Then, the mixing matrix $\textbf{H}$, which satisfies $[\textbf{R}_1,\textbf{R}_2]^T = \textbf{H}\times[\textbf{S}_1,\textbf{S}_2]^T$, can be expressed as Eq. \ref{eq:mixing_matrix}

\begin{equation}
\textbf{H}(t) = \frac{\lambda}{4\pi}
\begin{bmatrix}
\frac{G_{\textrm{tx1},1}(t)}{ \|\textbf{p}_{\textrm{tx1}}(t)-\textbf{p}_{\textrm{rx}}(t)\| } & \frac{G_{\textrm{tx2},1}(t)}{ \|\textbf{p}_{\textrm{tx2}}(t)-\textbf{p}_{\textrm{rx}}(t)\| } \\
\frac{G_{\textrm{tx1},2}(t)}{ \|\textbf{p}_{\textrm{tx1}}(t)-\textbf{p}_{\textrm{rx}}(t)\| } & \frac{G_{\textrm{tx2},2}(t)}{ \|\textbf{p}_{\textrm{tx2}}(t)-\textbf{p}_{\textrm{rx}}(t)\| } \\
\end{bmatrix}
\label{eq:mixing_matrix}
\end{equation}

\begin{figure}[ht!]
\centering
\includegraphics[width=.5\linewidth]{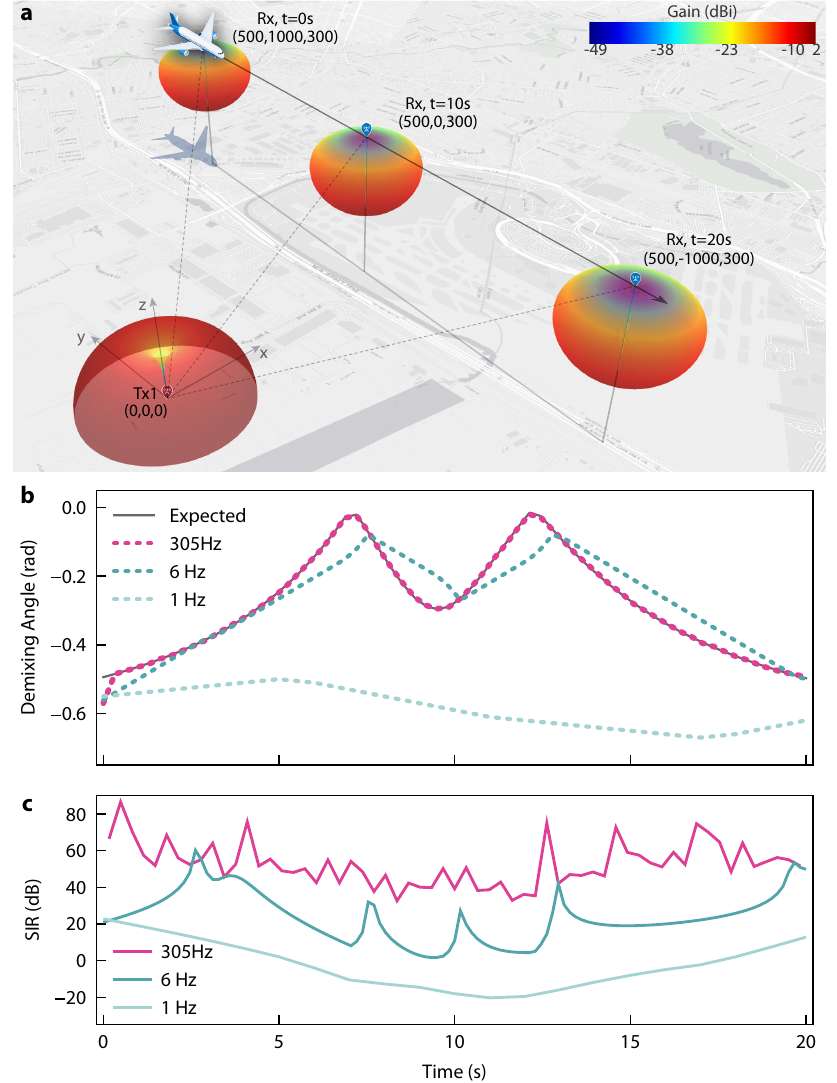}
\centering
\caption{\textbf{Dynamic interference simulation.} \textbf{a} Visualization of the simulated scene. A cellular tower, as the interference transmitter (Tx1), is located at the origin. All positions shown in the plot are in meters. An aircraft, as both the transmitter for SOI (Tx2) and the receiver (Rx), starts at position (300,1000,300) at $t=0$ s and moves at a steady height of 300 m with a constant speed of 100 $\textrm{m s}^{-1}$. The radiation pattern of Tx1 is shown in the plot centred at the transmitter near the origin. Tx1 is an omnidirectional dipole, which has vertical polarisation. The Rx is a 2x2 MIMO directional receiver with a boresight toward the z-axis ($(0,0,1)$). The Rx composes two orthogonal polarized antennas, aligning at $\textbf{e}_\textrm{rx,1}=(\sqrt{2}/2,-\sqrt{2}/2,0)$ and $\textbf{e}_\textrm{rx,2}=(-\sqrt{2}/2,\sqrt{2}/2,0)$. Due to symmetry, the radiation patterns of the two Rx antennas are identical. The Tx2 also share the same radiation pattern, but its polarisation slant is toward $\textbf{e}_\textrm{tx2}=(1,0,0)$. \textbf{b} Demixing angles for recovering Tx2 signal obtained by simulation of BSS algorithm. Grey curve represents the expected demixing angle, and dots in red, green, and light green are for the obtained demixing angles by a simulated BSS processor with update rates of 305 Hz, 6 Hz, and 1 Hz, respectively. \textbf{c} Signal-to-interference ratio (SIR) obtained under three simulated update rates. SIR is calculated by dividing the SOI power and the interference in the demixed output corresponding to the SOI. The output from the other channel is disregarded in this simulation.
}
\label{fig:simulation}
\end{figure}

In a simplified scenario of interference between a radar altimeter and a cellular network, as shown in Fig \ref{fig:simulation}a, an aircraft equipped with a 2x2 MIMO antenna (Rx, directional) can detect both the desired radar signal and unwanted interference from a nearby cellular tower (Tx1, omnidirectional). The radar signal is emitted downward from the aircraft and is received after a vertical round-trip between the aircraft and the ground. As the aircraft moves at a constant speed of $\textbf{v}=(0,100,0) \textrm{m s}^{-1}$, the distance and antenna gain of the cellular signal change, while those of the radar signal remain constant. In this simulation, we harness the polarisation selectivity of antennas to ensure that two mixtures possess distinct mixing ratios. In practice, we can exploit other antenna characteristics, such as uneven frequency responses and non-isotropic spatial gain, to achieve the desired variance in mixing ratios even if the SOI and interference do not differ regarding polarisation. By simulating this scenario for a period of 20 seconds, we obtain a time-variable mixing matrix and determine the ideal demixing weights as elements in one column of the inverse matrix $\textbf{H}^{-1}(t)$. Demixing weights that recover the radar signal (SOI) are represented as a vector, ($\textbf{V}=(w_1,w_2)$). This representation inherently mirrors the correlation coefficients utilized in the FastICA algorithm. This vector can be further delineated in terms of its amplitude ($a$) and angle ($\phi$). Considering that in practical systems, these demixing vectors are often scaled to their maximum to achieve the peak amplitude of the demixed outputs, it becomes evident that the demixing angle, $\phi$, is more informative. Hence, it is a practical one-scalar representation of the demixing vector.

With the mixing matrix obtained, we use it to test our BSS algorithm by applying it to two uncorrelated signals (regarded as the original SOI and the interference) at different update rates. Details of this simulation, including antenna specifications, generation of tested signals, and implementation of the BSS algorithm, can be found in Supplementary Fig. 4. The curves in Fig. \ref{fig:simulation}b. illustrate the obtained demixing angle for update rates of 305 Hz, 6 Hz, and 1 Hz. As previously demonstrated, BSS setups can suffer from laggy inter-device communications and low time-efficient statistical analysis, taking more than 1 second for a single iteration \cite{zhang2022broadband}. A BSS processor with poor time performance, represented by the green curves in Fig. \ref{fig:simulation}b and c, fails to keep up with changes in the demixing angle, resulting in a deterioration of signal-to-interference ratio (SIR) as low as -20 dB. On the other hand, a real-time operable BSS setup with a fast update rate of 305 Hz can significantly reduce error and enhance the SIR to at least 40 dB. This comparison demonstrates the significance of a higher update rate in achieving better performance when dealing with dynamic interference.

\subsection{Demonstration on radar altimetry}

\begin{figure}[ht!]
\centering
\includegraphics[width=.99\linewidth]{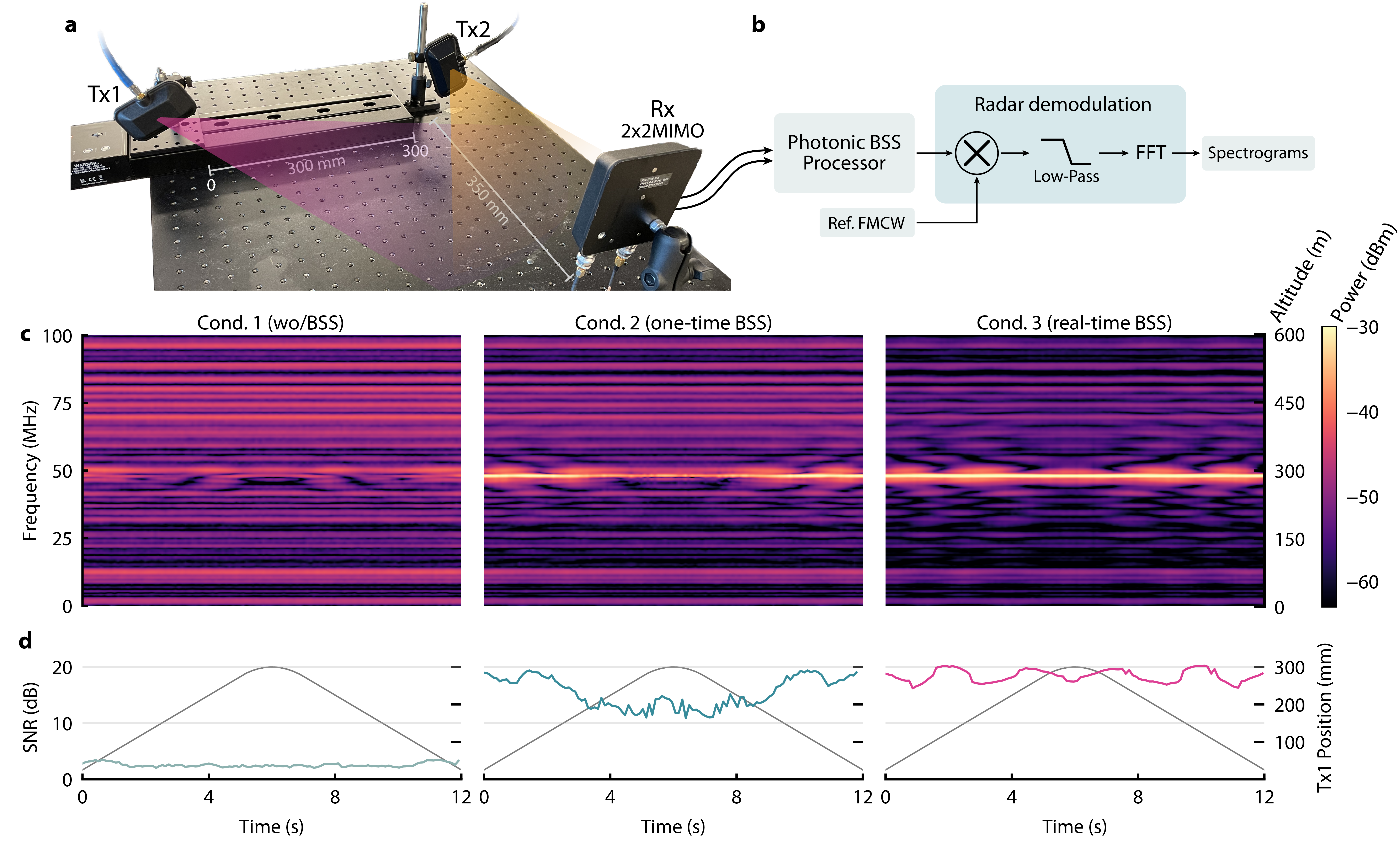}
\centering
\caption{\textbf{Experimental demonstration on radar altimeter.} \textbf{a} Experimental setup of dynamic interference environment. Tx1, Tx2, transmitters. Rx, receiver. MIMO, multiple-input and multiple-output. \textbf{b} Flow chart for demodulation of FMCW signal. \textbf{c} Spectrogram of demodulated FMCW signal for three tested conditions, calculated by Short-Time Fourier Transform on data sequences of 6.2 $\upmu$s. \textbf{d} Signal-to-noise ratios for the three tested conditions. Trajectories of antenna Tx1 are plotted in grey colour.
}
\label{fig:fmcw}
\end{figure}

In addition to simulations, we perform two experimental tests to demonstrate the effectiveness of the proposed photonic BSS processor in cancelling dynamic interference. The first demonstration focuses on eliminating interference between signals of a radar altimeter and a 5G cellular network. Radar altimeters utilise a frequency-modulated continuous wave (FMCW) signal format, with a chirp frequency ranging from 4.2 to 4.4 GHz \cite{raney1998delay}. Meanwhile, 5G cellular networks commonly use digital modulation formats, such as QPSK or 16-QAM, with multiple subcarriers following the orthogonal frequency division multiplexing (OFDM) mechanism \cite{farhang2016ofdm}. The licensed operating frequency band can vary in different countries and regions \cite{access2011user}. Though there is a guard band of about 200 MHz, the spurious emission of 5G signals can spectrally overlap with the radar signals. To demonstrate the worst-case scenario, we use a 16-QAM waveform centred at 4.3 GHz to represent the 5G interference. At the same time, the SOI is an FMCW format, continuously chirping from 4.2 to 4.4 GHz at every 4 $\upmu\textrm{s}$. These original signals are digitally designed, including baseband generation and up-conversion, then transmitted over the air using a multichannel arbitrary waveform generator (Keysight, M8196A) and transmitters.

As shown in Fig. \ref{fig:fmcw}a, the experimental implementation employs three antennas, including two transmitters (Southwest Antennas, 1004-022) and a 2x2 MIMO receiver (Southwest Antennas, 1055-395). The transmission distance is larger than 350 mm, meeting the far-field condition that requires at least 39 mm, as calculated by $2D^2/\lambda$, where $\lambda$ is the wavelength (0.75m) for a transmitted signal of 4 GHz and $D$ is the biggest antenna dimension (0.12 m). To create a dynamic interference scenario, we move one of the transmitters using a motorised translation stage (Thorlabs, LTS300) with a maximum speed of 50 $\textrm{m s}^{-1}$ and a range of 300 mm. Signals transmitted from the AWG have a power of 10 dBm (2 Vpp). The receiver has two outputs, each aligned to an orthogonal polarisation slant, resulting in mixtures with distinct mixing ratios. The received power for each output is approximately 2 dBm for this setup. The photonic processor receives these two mixtures as inputs ($\textbf{R}_1$ and $\textbf{R}_2$ in Fig. \ref{fig:schematic}a). At the same time, none of the channel state information, signal attributes, or any other prior knowledge is known or measured by the photonic processor.

In a real-world scenario, the transmitter and receiver for a radar altimeter system are co-located onboard, resulting in a round-trip for FMCW signals between aircraft and ground. Given the space of our laboratory, we generate the reflected signal by adding a time delay to the original FMCW signal and emitting it from one of the transmitters (Tx2) facing the receiver, thereby emulating the long propagation and reflection in actual radar altimetry. Mounted on the translation stage, another antenna (Tx1) transmits an interference 5G signal. On a typical optical table (0.6 m by 1.2 m), this configuration replicates, to a large extent, the scenario depicted in Fig. \ref{fig:simulation}a, where a flying trajectory is at a constant altitude of approximately 290 m and is located near a 5G cell tower. As the motorised stage moves reciprocally, the ($\|\textbf{p}_{\textrm{tx2}}(t)-\textbf{p}_{\textrm{rx}}(t)\|$) remains constant while the (($\|\textbf{p}_{\textrm{tx1}}(t)-\textbf{p}_{\textrm{rx}}(t)\|$)) sequentially decreases and increases.

Recovered signals from the photonic processor are analyzed by an oscilloscope (Tektronix, DPO73304sx), which also receives an original, non-delayed FMCW signal as the reference signal for demodulating altitude. After a pipeline of multiplication, low-pass filtering, and Fourier transformation (illustrated in Fig. \ref{fig:fmcw}b), generated spectrograms are depicted in Fig. \ref{fig:fmcw}c. Given that the delay between the two FMCW signals is constant at 0.96 $\upmu$s and the above-mentioned frequency sweeping range of 200MHz during a 4 $\upmu$s chirp time, the results should be a single-tone note around 48 MHz ($0.96\upmu\textrm{s}/4\upmu\textrm{s}\times200\textrm{MHz}$), indicating an altitude of 287.8 m. To demonstrate the efficacy of achieved real-time operability, tests are conducted under three conditions. The first condition is that demixing weights are fixed to the most positive weights such that $w_1 = w_2 = 1$ and no BSS algorithm is executed during the testing period. The second condition is that BSS is performed once when the moving antenna is in the home position, and then the BSS is disabled, and demixing weights are fixed for the remainder of the experiment. In the third condition, BSS remains active, and demixing weights are continuously updated throughout the duration of the test. In Fig. \ref{fig:fmcw}c, signal-to-noise ratios (SNR) of the demodulated tone signals under these conditions are depicted in light green, green, and pink, respectively. Only when BSS is continuously on duty (condition 3) can the SNR remain above 15 dB. The remaining two exhibit either indistinguishable interference signals (SNR $<$ 5) or excessive SNR variation (between 10 and 20 dB) that makes altitude measurement susceptible to error.

\subsection{Demonstration on 2.4 GHz band}

\begin{figure}[ht!]
\centering
\includegraphics[width=.99\linewidth]{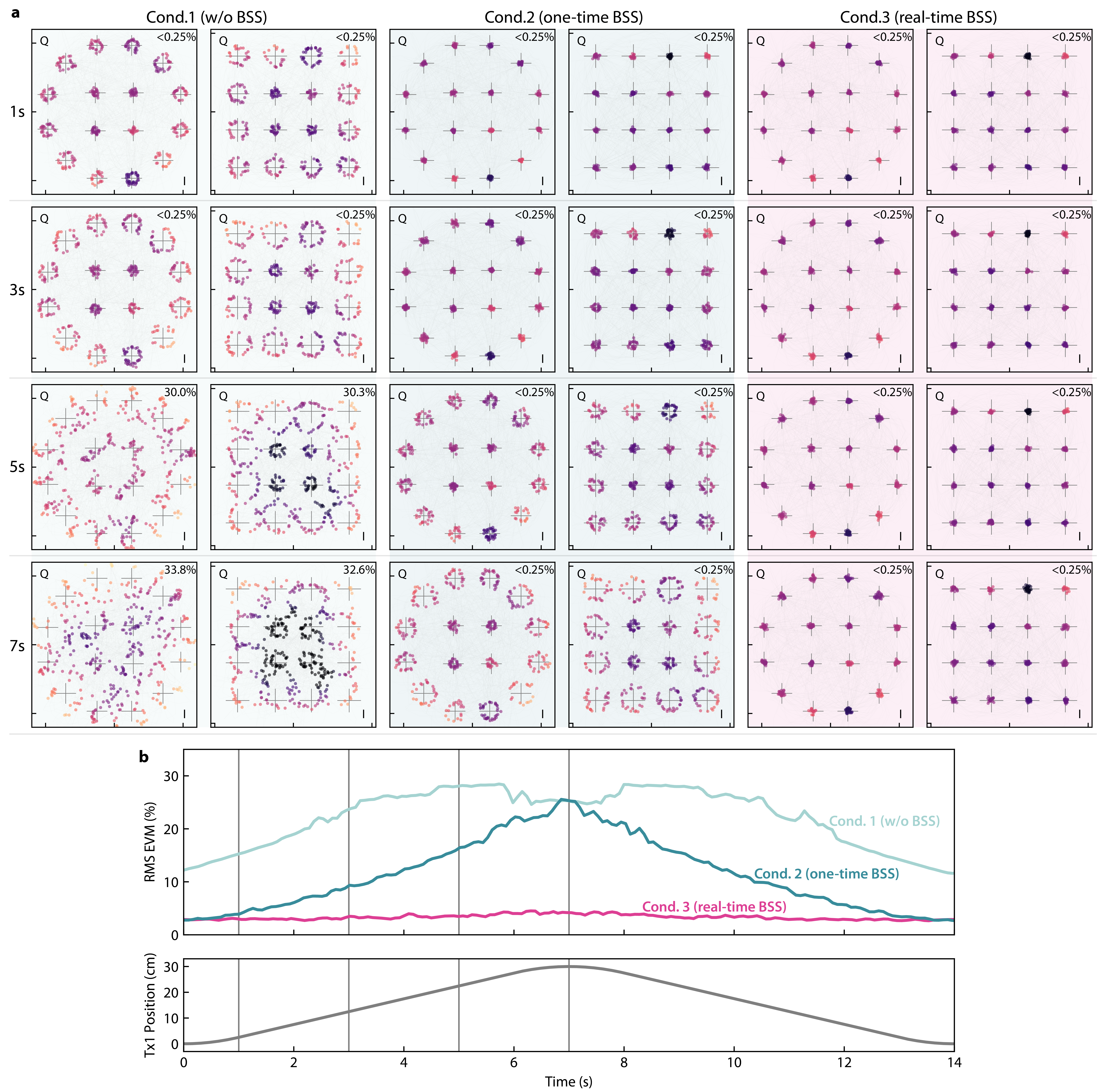}
\centering
\caption{\textbf{Experimental demonstration on 2.4 GHz band.} \textbf{a} Constellation diagrams of signals output by the photonic processor. Four diagrams in each column are for the same testing conditions, and each row corresponds to the same capture time (same antenna position). Horizontal axis represents the in-phase component (I), and vertical axis represents the quadrature component (Q). Bit error rate (BER) values are given at the top right corner of each diagram. Given the recorded number of symbols is 800, the lowest distinguishable BER is $0.125\%$. \textbf{b} EVM variations and position of the moving transmitter during the test period. Pink, green, and light green curves are the averaged EVMs of two recovered signals obtained under conditions 1, 2, and 3, respectively. Grey curve in the bottom panel is the transmitter position. During the test, Tx1 completes one round trip for about 14 s.
}
\label{fig:constellation}
\end{figure}

The second demonstration, shown in Fig. \ref{fig:constellation}, investigates interference in the notoriously congested 2.4 GHz band. Due to license-free convenience and demand for the Internet of Things, this band is filled with numerous wireless signals of standard services (WiFi, Bluetooth) and proprietary protocols (image transmission, remote control), many of which are extensively used for mobile objects such as drones, ground vehicles, and mobile phones. To accommodate the desired frequencies, we adopt the same configuration shown in Fig. \ref{fig:fmcw}a but modify it with two new transmitters (Southwest Antennas, 1004-034) and receivers (Southwest Antennas, 1055-368). Two transmitted signals have uncorrelated content (two sets of random bits) and are generated using different digital modulation formats (16-QAM and 16-APSK). But they share the same carrier frequency of 2.4 GHz and symbol rate of 400 MHz, posing a challenging problem that spectral filtering techniques cannot solve. To exploit the capability of this photonic processor, for this time, we use all two signal pathways in parallel and aim to recover both the original signals simultaneously.

To evaluate the performance, outputs are recorded and demodulated using an oscilloscope with built-in software (Tektronix, SignalVu). Results are presented through constellation diagrams and error vector magnitudes (EVM) in  Fig. \ref{fig:constellation}, which demonstrates the effectiveness of the BSS algorithm in restoring the desired SOI from a mixture of signals. Each frame has a recording length of 2 microseconds and contains 800 symbols, with a symbol rate of 400 MSPS. Comparisons between results from different testing conditions reveal that the BSS algorithm can significantly lower EVM by up to eight times (from $26\%$ to $3\%$ at $t=7$ s). It can eliminate errors in transmitted messages, reducing the bit error ratio (BER) from higher than $30\%$ to less than $0.25\%$. Furthermore, when comparing the second and third conditions, it is clear that the continuous BSS operation can track variable mixing ratios and keep EVM at a constant low level (less than $4\%$) while preventing bit errors (BER $<0.25\%$) throughout the entire testing period. These results exemplify the efficacy of the proposed real-time photonic BSS system in recovering SOI in dynamic interference environments.

\section{Conclusion}

This study initiates the exploration of instantaneous adaptability in PICs, demonstrating real-time online learning and adjustment of photonic weights with a millisecond updating rate, which ushers in a new paradigm for operating photonic devices. This work achieves a notably reduced latency by fully integrated on-chip signal pathways, ensuring preparedness for future technological demands of increasingly high carrier frequencies and data rates. Two experimental tests prove this setup can address challenging RF interference problems with dynamic mixing ratios. This work carries forward the unmatched advantages of photonic signal processing to solving complex real-world tasks in a field-ready form factor. 

In this work, our implemented MRR weight banks primarily address amplitude mismatch but do not provide time delay for phase matching compensation. Future improvements could be to incorporate on-chip time delay elements like racetrack delay lines \cite{melloni2010tunable} and binary switched delay lines to address phase errors and handle complex mixing matrices. Additionally, we aim to enhance the link loss by optimising the design of the BPDs, the TIAs, and the wire-bonding lengths. Also can be improved is the processing bandwidth, which is primarily limited by the EO modulators, specifically the PN junction-typed MRRs. While they should theoretically achieve bandwidth in tens of GHz \cite{zhang2020200}, our results are constrained by the inductance of extended bonding wires and sub-optimal on-chip components. For greater bandwidth at the expense of scalability, Mach-Zehnder modulators are also an option. Once EO modulation limits are addressed, the photodetector, currently benchmarked at around 16 GHz \cite{hai201316}, becomes the next focus for enhancement.

The WDM compatibility inherent to our MRR-based design provides a tangible avenue for system scalability, allowing for more intricate computational tasks beyond the currently demonstrated 2x2 matrix operations. Yet tapping into larger-scaled applications like massive MIMO systems brings its own challenges, such as orchestrating resonance frequencies for a vast array of MRRs to avert spectral overlap and the need to source lasers for a plethora of wavelengths. Potential resolutions include replacing MRRs with photonic crystal cavities to ensure a broader FSR and utilizing comb lasers as a compact solution for multi-wavelength generation.

As we continue refining form factors, enhancing performance, and augmenting online adaptiveness, we foresee a promising future wherein photonic processors cater to a broader range of demanding applications. This includes but is not limited to model predictive control \cite{huang2022prospects,de2019machine} and neuromorphic computing. With these advancements, the potential impact and utility of photonic processors have become increasingly significant.

\section{Methods}
\subsection{Microring weight bank}
Laser lights that enter a weight bank from the input port (In) are filtered by corresponding MRRs with adjacent resonance frequencies, where optical power is separated into the drop port and the thru port (as depicted in Fig. \ref{fig:schematic}e). A balanced photodetector sums up the difference between these two ports, giving weighted addition of all the input signals. Tuning the resonance of each MRR can vary the portion of the optical power that goes into each of the two ports, namely, controlling the weights. Noting that the non-uniform transmission profile of the MRR could introduce non-linearity during  EO modulation and weighting, resulting in a Spurious Free Dynamic Range (SFDR) reduction from 25.5 dBc to 11.5 dBc for a 4 GHz single-tone signal. Details of this SFDR measurement and discussion on potential improvement can be found in Supplementary Information. The photonic chip was fabricated on a silicon-on-insulator wafer with a silicon thickness of 220 nm and a buried oxide thickness of 2 µm. The waveguide is 500 nm wide. The weight bank consists of four MRRs (radius around $\textrm{r}=8\;\upmu\textrm{m}$) coupled with two bus waveguides in an add/drop configuration, and two among the four were used in experiments, which have resonance frequencies of around 193405 GHz (1550.07 nm) and 193435 GHz (1549.83 nm), respectively, at 25-degree Celsius. The ring radii reveal slight difference ($\Delta\textrm{r}=12\;\textrm{nm}$) to avoid resonance collision. The gap between the ring and bus waveguide is 200 nm, yielding a Q factor of about 6,000. Associated circular metal heaters can thermally tune the weights of each MRR. Metal vias and traces were deposited to connect the heater contacts of the MRR weight bank to electrical metal pads, and the heater resistance is around 2000 Ohms.

\subsection{Photonic chip packaging}
The PIC, together with two TIAs and a few decoupling capacitors, are mounted directly onto the top PCB with silver epoxy, and they are wire-boned to pads on the PCB for electrical connection. The PCB uses Rogers 4003C as the substrate material to support high-frequency signals at a low loss. RF signal input and output the PCB via SMP connectors and are routed to the PIC through impedance-controlled traces. Optical coupling is through a glued fibre array, which has a polished angle of 41 degrees to allow working with on-chip grating couplers of an 8-degree incident angle. The coupling loss is tested to about 7 dB per single trip between the fibre array and the chip. Connected via the ribbon cable, the bottom PCB offers tuning currents for MRR weight banks and biasing voltages for MRR modulators and BPDs. This bottom populates two 5-channel current output DACs (LTC2662, Analog Devices) and one 8-channel voltage DAC (LTC2686, Analog Devices) are populated. Also populated are several low-noise linear power supply chips for the demanded power rails (+5V, +10V, -10V). This bottom board is connected to the FPGA board and can be programmed via a high-speed SPI protocol. Besides, a thermal-electrical cooler is a bandwidth between the two PCBs and a thermistor mounted on the top board, which is utilized by a temperature controller for thermal stability.

\subsection{FPGA-based dithering control}
The dithering control method can compensate for vulnerability to environmental fluctuations, bringing about 9-bit high weighting accuracy (up to 9-bit) for sensitive photonic devices, such as MRR and MZM. In this work, we implemented complete dithering control using the co-packaged FPGA. Two built-in digital-to-analogue converters (DAC) generated two signal-toned signals at 10 MHz and 15 MHz, dithering the two received mixtures via a duplexer at the two MZMs. After being processed by the photonic chip, the output is digitized by the built-in ADC of the FPGA. Then, besides the rest BSS pipeline, we use extra programmable logic in the FPGA to perform I-Q demodulation at the two dithering frequencies. Since the dithering signals share the same pathway as the actual signals, they are processed with identical weights by the photonic processor. Therefore, the resulting amplitudes of the I-Q vectors are proportional to the weights. Thanks to the ADC and DACs being operated by synchronized clocks, the phases of the I-Q vectors are stable and can determine the sign. With this, the actual weights are measured in every iteration. These measurements are accessible to the high-level BSS program running in the ARM processor core, allowing adjusting the MRR currents for accurate weighting.

\subsection{Real-time BSS algorithm}
Guided by the centre limit theorem, BSS pursue the maximal non-gaussianity to search the demixing target weights that can recover original signals from their mixtures. Previous attempts \cite{zhang2022broadband,ma2020blind} use optimization algorithms, such as NM \cite{nelder1965simplex} that work for a problem with a static objective function (demixing weights versus output kurtosis) and do not react to moving optimal points. To accommodate dynamic RF interference, we revise the standard NM method. NM denotes the searching range as simplex, which expresses the demixing weights in this work. Conventional NM searching does not limit simplex with a minimal size, which leads to a monotonous tendency of diminishing the simplex size when approaching an optimal point. The resulting simplex vertices are too closely located and fail to perceive and respond to a rapid shift in the objective function. Awareness of this shift needs to maintain a minimal simplex size, but a trade-off exists that an enlarged simplex can lead to deviation from the optimal, revealing increased errors. In this work, we revised the NM setting a lower bound of $\Delta w = 0.008$ for the simplex size, namely the weight searching range, which successfully deals with all the dynamic RF interference problems and can maintain less than $0.4\%$ ($0.008/2\%$) per cent error.

\section{Data Availability}
All data used in this study are available from the corresponding authors upon request.

\section{Code Availability}
All codes used in this study are available from the corresponding authors upon request.

\backmatter






\bibliography{ref}

\section{Acknowledgments}

This research is supported by the National Science Foundation (NSF) (ECCS-2128616 and ECCS-1642962 to P.R.P.), the Office of Naval Research (ONR) (N00014-18-1-2297 and N00014-20-1-2664 P.R.P.), and the Defense Advanced Research Projects Agency (HR00111990049 to P.R.P.). The devices were fabricated at the Advanced Micro Foundry (AMF) in Singapore through the support of CMC Microsystems. A. Tait and B. J. Shastri acknowledge support from the Natural Sciences and Engineering Research Council of Canada (NSERC). S. Bilodeau acknowledges funding from the Fonds de recherche du Québec - Nature et technologies.

\section{Author contributions}
W.Z., J.C.L., and T.F.L. conceived the ideas. W.Z., J.C.L., and J.Z. implemented the experimental setup, designed the experiment, and conducted the experimental measurements. W.Z. and L.S.H. analyzed the results. T.F.L. and S.B. designed the silicon photonic chip. T.F.L., A.T., and B.J.S. provided theoretical support. S.B. performed the chip packaging. W.Z., J.C.L., T.F.L., S.B., L.H., and B.J.S. wrote the manuscript. P.R.P. supervised the research and contributed to the general concept and interpretation of the results. All the authors discussed the data and contributed to the manuscript.

\section{Competing interests}
The authors declare no competing interests.

\end{document}